\documentstyle[12pt,
         epsfig]{article}
\def\beq{\begin{equation}} 
\def\eeq{\end{equation}} 
\def\bea{\begin{eqnarray}} 
\def\eea{\end{eqnarray}} 
\def\bq{\begin{quote}} 
\def\eq{\end{quote}}
\def\ra{\rightarrow}

\def\bq{\begin{quote}} 
\def\eq{\end{quote}}
\def\ra{\rightarrow}

\def\ltap{\ \raisebox{-.4ex}{\rlap{$\sim$}} \raisebox{.4ex}{$<$}\ }

\newcommand{\labgraph}[2]{%
   \begin{picture}(3,4)
       \put(0,0){\makebox(3,2)%
{\centering\epsfxsize=3in\leavevmode\epsffile{#2.ps}}}
       \put(0,1){\makebox(0,0)[l]{(#1)}}
   \end{picture}}
\def\sublabel#1#2{\@bsphack\if@filesw {\let\thepage\relax
   \def\protect{\noexpand\noexpand\noexpand}%
   \edef\@tempa{\write\@auxout{\string
      \newlabel{#1}{{\@currentlabel#2}{\thepage}}}}%
   \expandafter}\@tempa
   \if@nobreak \ifvmode\nobreak\fi\fi\fi\@esphack}
\newcommand{\threegraphs}[3]{%
   \unitlength=1in
   \begin{center}
     \begin{picture}(6,7)
       \put(0,4){\labgraph{a}{#1}}
       \put(3,4){\labgraph{b}{#2}}
       \put(1.5,0.5){\labgraph{c}{#3}}
     \end{picture}
   \end{center}}

\begin{document}
\baselineskip 18pt
\newcommand{\sheptitle}
{Hybrid MSW + VO Solution of the Solar Neutrino Problem\\
 in String-Motivated Unified Theories
}
\newcommand{\shepauthor}
{B. C. Allanach$^1$, G. K. Leontaris$^{2,3}$ and  S. T. Petcov$^{4,5}$}
\newcommand{\shepaddress}
{1. Rutherford Appleton Laboratory, Chilton, Didcot, Oxon. OX11 0QX, \\United
  Kingdom \\
  2. CERN Theory Division, 1211 Geneva 23, Switzerland\\
 3. Physics Department,  University of Ionnanina, Ioannina, GR-45110, Greece \\
  4. Scuola Internazionale Superiore di Studi Avanzati (SISSA), and 
Istituto Nazionale di Fisica Nucleare, I-34013, Trieste,  Italy \\
  5. Institute of Nuclear Research and Nuclear Energy, Bulgarian Academy of
Sciences, 1784 Sofia, Bulgaria}

\newcommand{\shepabstract}
{It is shown that the hybrid MSW + VO solution of the solar neutrino problem,
according to which the solar $\nu_e$ undergo matter-enhanced transitions
into $\nu_{\mu, \tau}$ in the Sun followed by long wave length
($\sim 1.5\times 10^{8}~km$) $\nu_e \leftrightarrow \nu_{\mu, \tau}$
oscillations in vacuum between the Sun and the Earth, can occur naturally
in string-motivated grand unified theories. We consider the supersymmetric version
of a string-type $SU(4)\otimes SU(2)_{L}\otimes SU(2)_{R}$ theory with
$U(1)_{X}$ family symmetry, which was shown to successfully describe 
the charged fermion masses and the quark mixing, and extend the earlier fermion mass
analysis to the neutrino sector. 
We show that the four oscillation parameters
$\Delta {m_{31}^2}$, $\Delta {m_{21}^2}$ and $\sin^2 2\theta_{12}$,
$\sin^2 2\theta_{13}$, characterising the combined matter-enhanced transitions 
and vacuum oscillations of the solar $\nu_e$, naturally get values in the
ranges 
of the hybrid MSW + VO solutions found recently.
}

\begin{titlepage}
\begin{flushright}
CERN-TH-97-347\\
RAL-TR-97-062\\
IOA-TH-97-017\\
Ref. SISSA 155/97/EP\\
hep-ph/9712446\\
\end{flushright}
\vspace{.1in}
\begin{center}
{\large{\bf \sheptitle}}
\bigskip \\ \shepauthor \\
\vspace{\baselineskip} 
\shepaddress \\
\mbox{} \\ \vspace{.2in}
{\bf Abstract} \bigskip 
\end{center} 
\setcounter{page}{0}
\shepabstract
\end{titlepage}
\section{Introduction}
\vskip -0.2truecm
 In the present article we show that the hybrid matter-enhanced 
  transitions +
vacuum oscillations solutions of the solar neutrino problem, possible 
in the case of three flavour neutrino mixing and  
found in ref. \cite{LP97}, can appear naturally in a 
class of string-type unified theories.
As is well-known, the solar neutrino observations provide 
strong evidences for the 
existence of non-zero neutrino masses and lepton mixing.
The signals measured in all six experiments which have obtained data
on the solar neutrino flux so far, Homestake, Kamiokande II and III, SAGE, GALLEX and 
Super-Kamiokande, are considerably smaller than the 
signals predicted by the standard solar models. The case for a significant
solar neutrino deficit was reinforced recently
\cite{HELIOSEIS1} 
with the publication of new more precise helioseismological data 
(see \cite{HELIOSEIS1} 
for the relevant references). 
These data made possible to perform a number of critical tests of the solar
models, including the models which have been constructed with the
purpose of solving the solar neutrino problem without invoking 
``unconventional'' neutrino behaviour
(matter-enhanced transitions, vacuum oscillations, etc.).
The analyses completed so far showed \cite{HELIOSEIS1}
that only the standard
solar models which include the diffusion of heavy elements
in the solar interior are compatible with the most
recent helioseismological observations, and that none of these models
can describe the current solar neutrino data.
The indicated results make the possibility of an
``unconventional'' behaviour of the solar $\nu_e$
on their way from the central part of the Sun to the Earth surface   
look at present more plausible than ever in the past.
However, the crucial unambiguous experimental proof for such a behaviour
is still lacking and it is hoped that the presently operating and the future
solar neutrino experiments will be able to determine the true cause of the solar
neutrino problem.
   
      The hybrid resonance transitions + vacuum oscillations (MSW + VO) solution
is based on the assumption  
that the solar $\nu_e$ undergo matter-enhanced transitions
into $\nu_{\mu}$ and/or $\nu_{\tau}$ 
when they propagate from the central part to the surface of the Sun and that 
these 
transitions are followed by long wave length ($\sim 1.5 \times 10^{8}~km$) 
vacuum oscillations
of the $\nu_e$ and/or $\nu_{\mu(\tau)}$ when the
flavour neutrinos travel from the surface of the Sun to the Earth.
The solution of interest is a genuine three-flavour-neutrino mixing solution 
since it is possible only if the three flavour neutrinos mix in vacuum:   
\begin{equation}
|\nu_l>~= \sum_{k=1}^3 U_{lk}^* |\nu_k>, 
\hspace{1cm} l=e,\mu,\tau,
\end{equation}
\noindent where $|\nu_l>$ is the state vector of
the (left-handed) flavour neutrino $\nu_l$ having a definite momentum
($\overrightarrow{p}$), $|\nu_k>$ is the state vector of a neutrino
$\nu_k$ possessing a definite mass $m_k$ and a definite momentum ($\overrightarrow{p}$),
$m_k \neq m_j$, $k \neq j = 1,2,3$, $m_1 < m_2 <  m_3$\footnote{It is always
  possible to choose
$m_1 < m_2 <m_3$ without loss of generality
and we will work with this convention in what follows.}, 
and $U$ is a $3 \times 3$
unitary matrix -- the lepton mixing matrix. The (average) 
solar $\nu_e$ survival probability,  
$\bar{P}(\nu_e \rightarrow \nu_{e})$, 
which is a basic quantity in the analysis of the solar neutrino data 
in the case of interest and for which a simple analytic 
expression was derived in \cite{LP97},
depends on four neutrino mass and mixing parameters: 
two basically responsible for the matter-enhanced
transitions and another two responsible for the
long wave length vacuum oscillations. In view of our further discussion
we will assume these parameters to be 
respectively $\Delta m^2_{31}$, $\sin^22\theta_{13}$ and
$\Delta m^2_{21}$, $\sin^22\theta_{12}$, where as usual
$\Delta m^2_{ij} = m^2_i-m^2_j$, and   
\begin{equation}
\cos2\theta_{13} = 1 - 2|U_{e3}|^2 > 0,~~\sin2\theta_{13} = 2|U_{e3}|\sqrt{1
  - |U_{e3}|^2}~> 0,
\end{equation}
\begin{equation}
\sin^22\theta_{12} = 4 {{|U_{e1}|^2~|U_{e2}|^2}\over {(|U_{e1}|^2 +
|U_{e2}|^2)^2}}
\end{equation}
\noindent 
As was shown in ref. \cite{LP97} on the basis of an analysis of the recent 
solar neutrino data, 
the MSW + VO solution is possible for 
\begin{equation}
4.0 \times 10^{-12} eV^2 \ltap \Delta m^2_{21}
      \ltap 5.0\times 10^{-10}eV^2,
\end{equation}
\begin{equation}
 0.15 \ltap \sin^22\theta_{12} \leq 1.0,
\end{equation}
\noindent and
\begin{equation}
5.0 \times 10^{-6}eV^2 \ltap \Delta m^2_{31}
        \ltap 1.5\times 10^{-4}eV^2,
\end{equation}
\begin{equation}  
3.1\times 10^{-4} \ltap \sin^22\theta_{13} \ltap 0.5.
\end{equation}
\noindent More specifically, in \cite{LP97} 
five different sub-regions in the region of the space of parameters, defined
by eqs. (2) - (7), corresponding to five different types of hybrid 
MSW + VO solutions, A, B, C, D, and E,  have been identified
(see Figs.~\ref{fig:LP}a-\ref{fig:LP}c).
Let us add that, what actually specifies the solutions
$A$, $B$, $C$, $D$ and $E$ 
and makes them physically very
different is the distinct way
the various solar neutrino flux components
($pp$, $^{7}$Be, $^{8}$B, etc.) are affected
by the matter-enhanced transitions and the vacuum oscillations
in the case of each particular solution (see \cite{LP97} for further details).

  As it follows from eqs. (2) - (7), the combined MSW + VO solution
requires rather small neutrino mass squared differences and very specific 
mixing in the lepton sector and ratio of $\Delta m^2_{21}$ and $\Delta m^2_{31}$. 
In the context of GUT's, the light
neutrino masses can be generated, as is well-known, by the see-saw mechanism: 
$m_{\nu}\sim m^2_Q/M_{GUT}$,
where $m_Q$ is the up quark mass and $M_{GUT}$ is the unification
scale. Moreover, making use of simple discrete or continuous
$U(1)$ symmetries at the unification scale, specific textures for 
the fermion mass matrices can be constructed with a minimum number of
parameters which leads, in particular, to rather specific predictions
for the mixing in the lepton sector. 
Effective unified or partially unified GUTs which 
incorporate naturally the indicated properties
can be built within the string theories. 
In the free fermionic formulation
 for example, a number of
such models has been constructed \cite{aehn,alr,af}. A typical feature of the
latter is that in addition
to the gauge group containing the SU(3)$\otimes$SU(2)$_L\otimes$U(1)$_Y$
as a subgroup, a number of surplus $U(1)$ symmetry factors which
distinguish between the three fermion families appear. However, models
based on these constructions  do not contain Higgs fields in the
adjoint or higher dimensional representations. As a consequence, the 
unification symmetry of the
traditional GUT's like SU(5) and SO(10), cannot be broken down
to the symmetry of the Standard Theory. 

   Attempts to overcome this difficulty led to
constructions of partially unified 
GUT's in which only small-dimensional Higgs representations are needed to break
the unification symmetry \cite{aehn,alr}.
A partially unified theory which fulfils all 
basic requirements for a string-type GUT
\cite{alr}, is  based on the Pati--Salam \cite{PS} gauge symmetry group 
SU(4)$\otimes$O(4). The SU(4)$\otimes$O(4)
unification symmetry can be broken down \cite{alr}
to the standard model gauge symmetry without using adjoint or any higher
dimensional Higgs representations. The model possesses a number 
of attractive features.
Colour triplets and Higgs doublets arise in different
representations. Thus, it is free from doublet--triplet splitting
complications as the triplets become massive from simple trilinear
couplings. There are no dangerous proton decay mediating  gauge
bosons. As a consequence, the SU(4)$\otimes$O(4) breaking scale $M_{GUT}$ can in
principle be lower than the 
 symmetry breaking scale of other unified theories, (e.g., the SO(10) 
theory).  In fact, the basic constraints on $M_{GUT}$ are the low energy
measurements of the strong coupling constant 
$\alpha_s$ and of $\sin^2\theta_W$. Thus, in
the case when only the minimal supersymmetric standard model (MSSM)
  spectrum appears below $M_{GUT}$ these parameters
are correctly predicted if $M_{GUT}\sim 10^{16}GeV$.
{}Furthermore, a recent 
non-renormalisable operator analysis \cite{AKLL} of the model's 
supersymmetric version has revealed quite remarkable  features 
of the charged fermion mass matrices. 

  In this work we demonstrate that the hybrid MSW +
VO scenario, shown recently
to provide an alternative solution of the solar neutrino 
problem \cite{LP97} (see also, e.g., \cite{Panta91}), 
can be naturally implemented in the string-type 
partially unified SU(4)$\otimes$O(4) theory.
Our analysis will be performed in the context of the supersymmetric
version of the theory with the SU(4)$\otimes$O(4) gauge symmetry
augmented by a $U(1)$ family symmetry.
\section{The SU(4)$\otimes$O(4) Model \label{model}}
Here we briefly summarise the basic features of the model which are relevant
for our analysis \cite{AKLL}.  The gauge group is
SU(4)$\otimes$O(4)$\otimes$U(1)$_X$,  
or equivalently its isomorphic (enhanced) Pati-Salam symmetry
\begin{equation}
\mbox{SU(4)}\otimes \mbox{SU(2)}_L \otimes \mbox{SU(2)}_R
\otimes \mbox{U(1)}_X, \label{422}
\end{equation}
where U(1)$_X$ is a gauged family symmetry of the same type as discussed in
ref.~\cite{familysym}.
The left- and right-handed fermion fields are accommodated in the following
representations,
\begin{equation}
{F^i}^{\alpha a}=(4,2,1)=
\left(\begin{array}{cccc}
u^R & u^B & u^G & \nu \\ d^R & d^B & d^G & e^-
\end{array} \right)^i\label{LH}
\end{equation}
\begin{equation}
{\bar{F}}_{x \alpha}^i=(\bar{4},1,\bar{2})=
\left(\begin{array}{cccc}
\bar{d}^R & \bar{d}^B & \bar{d}^G & e^+  \\
\bar{u}^R & \bar{u}^B & \bar{u}^G & \bar{\nu}
\end{array} \right)^i\label{RH}
\end{equation}
where $\alpha=1,\ldots ,4$ is an SU(4) index, $a,x=1,2$ are
SU(2)$_{L,R}$ indices, and $i=1,2,3$ is a family index.  
The U(1)$_X$ family symmetry is broken spontaneously
below the string scale ${M_S}$ by the
superfields $\theta, \bar{\theta}$, both singlets under the 
$\mbox{SU(4)}\otimes \mbox{SU(2)}_L \otimes \mbox{SU(2)}_R$ symmetry,
but with U(1)$_X$ charges $+1,-1$ respectively.
This breaking cancels anomalous D-terms provided the vacuum expectation values
of $\theta$ and $\bar{\theta}$ satisfy~\footnote{Note that in order
  to avoid anomalies of the fermionic components and preserve flatness
conditions of the scalar VEVs, we introduce always $R+\bar R$
representations.} \cite{AKLL}
\begin{equation}
\langle \theta \rangle \ \sim \langle \bar{\theta} \rangle \sim
(0.1-0.3) {M_S}. \label{scal}
\end{equation}
The Higgs fields are contained in the following representations,
\begin{equation}
h_{a}^x=(1,\bar{2},2)=
\left(\begin{array}{cc}
  {h_2}^+ & {h_1}^0 \\ {h_2}^0 & {h_1}^- \\
\end{array} \right) \label{h}
\end{equation}
(where $h_1$ and $h_2$ are the low energy Higgs superfields associated
with the MSSM.) The two heavy Higgs representations are
\begin{equation}
{H}^{\alpha b}=(4,1,2)=
\left(\begin{array}{cccc}
u_H^R & u_H^B & u_H^G & \nu_H \\ d_H^R & d_H^B & d_H^G & e_H^-
\end{array} \right) \label{H}
\end{equation}
and
\begin{equation}
{\bar{H}}_{\alpha x}=(\bar{4},1,\bar{2})=
\left(\begin{array}{cccc}
\bar{d}_H^R & \bar{d}_H^B & \bar{d}_H^G & e_H^+ \\
\bar{u}_H^R & \bar{u}_H^B & \bar{u}_H^G & \bar{\nu}_H
\end{array} \right). \label{barH}
\end{equation}
At a scale $M_{GUT} \sim 10^{16}$ GeV, these Higgs fields are
assumed to develop VEVs,
$\langle H\rangle =\langle \tilde{\nu}_H\rangle \sim M_{GUT}, \ \ 
\langle \bar{H}\rangle =\langle \tilde{\bar{\nu}}_H\rangle \sim M_{GUT}$,
leading to the symmetry breaking at $M_{GUT}$
\begin{equation}
\mbox{SU(4)}\otimes \mbox{SU(2)}_L \otimes \mbox{SU(2)}_R
\longrightarrow
\mbox{SU(3)}_C \otimes \mbox{SU(2)}_L \otimes \mbox{U(1)}_Y
\label{422to321}
\end{equation}
in the usual notation.  

  Under the symmetry breaking eq. (\ref{422to321}),
the bidoublet Higgs field $h$ in eq. (\ref{h}) splits into two
Higgs doublets $h_1$, $h_2$ whose neutral components subsequently
develop weak scale VEVs,
\begin{equation}
\langle h_1^0\rangle =v_1, \ \ \langle h_2^0\rangle =v_2 \label{vevs1}
\end{equation}
with $\tan \beta \equiv v_2/v_1$.

In addition to the Higgs fields in eqs.~(\ref{H}) and (\ref{barH}) 
one introduces also an SU(4) sextet field $D=(6,1,1)$.
Assuming now a $Z_2$ invariance with respect to 
$H (\bar{H})\ra (-1)\times H (\bar{H})$, the tree level mass terms of
the superpotential of the model read 
\begin{equation}
W =\lambda^{ij}_1\bar{F}_i F_j h
+\lambda_2HHD+\lambda_3\bar{H}\bar{H}D+\mu hh \label{W}
\end{equation}
Note that we have banned terms which might lead to
unacceptably large neutrino-higgsino mixing~\cite{fp}.
Additional terms not included in eq. (\ref{W}) may be forbidden by imposing 
suitable discrete or continuous symmetries, the details of which
need not
concern us here.  The $D$ field carries colour and therefore 
does not develop  a VEV, but the terms in eq. (\ref{W})
$HHD$ and $\bar{H} \bar{H}D$ combine the  colour triplet parts of $H$, $\bar{H}$
and $D$ into acceptable GUT-scale mass terms. When the $H$ fields
attain their VEVs at $M_{GUT}\sim10^{16}$ GeV, the superpotential of eq.~(\ref{W})
reduces to that of the MSSM augmented by neutrino masses. Below $M_{GUT}$ the 
part of the superpotential involving matter superfields is just
\begin{equation}
W =\lambda^{ij}_U\bar{U}_iQ_jh_2+\lambda^{ij}_D\bar{D}_iQ_jh_1
+\lambda^{ij}_E\bar{E}_iL_jh_1+ \lambda^{ij}_N\bar{\nu}_iL_jh_2 + \ldots
\label{MSSMmatter}
\end{equation}
 where $Q_j$ and $L_j$ are the quark and lepton $SU(2)_{L}$ doublet
superfields and $\bar{U}_i$, $\bar{D}_i$, $\bar{E}_i$ and
$\bar{\nu}_i$ are the charge 2/3 and (-1/3) quark,
 the charged lepton and the
neutrino $SU(2)_{L}$ singlet superfields. 
The Yukawa couplings in eq.~(\ref{MSSMmatter}) satisfy the
boundary conditions
\begin{equation}   
\lambda^{ij}_1 (M_{GUT}) \equiv \lambda^{ij}_U(M_{GUT}) 
= \lambda^{ij}_D (M_{GUT})=
\lambda^{ij}_E(M_{GUT}) = \lambda^{ij}_N(M_{GUT}).
 \label{boundary}
\end{equation}
Thus, eq.~(\ref{boundary}) retains the successful relation
$m_{\tau}=m_b$ at $M_{GUT}$. 

 The boundary conditions listed in eq.~(\ref{boundary}) lead to
unacceptable mass relations for the light two families. In addition, the
large family hierarchy in the Yukawa couplings appears to be unnatural
since one would naively expect the dimensionless couplings all to be  
of the same order. This leads us to the conclusion that the
$\lambda^{ij}_1$ in eq.~(\ref{W}) may not originate from the usual
renormalisable tree level dimensionless coupling.  The $U(1)_X$
symmetry will allow a  
renormalisable Yukawa coupling in the 33-term only and generate the
rest of the effective Yukawa couplings by non-renormalisable operators
that are suppressed by some higher mass scale. This suppression
provides an explanation for the observed fermion mass hierarchy
and occurs as powers of two expansion parameters $\epsilon$ and 
$\delta $ defined as follows
\begin{equation}
 \epsilon \equiv \langle \theta\rangle /{M_S} \sim
\langle \bar{\theta}\rangle /{M_S} \sim M_X/M_S
,\,\,\,  \delta \equiv \langle H\rangle 
 \langle \bar{H}\rangle  / M_Y^2 = M_{GUT}^2 / M_Y^2
\end{equation}
where $M_X \sim \langle \theta\rangle \sim \langle \bar{\theta}\rangle$ 
is the $U(1)_X$ symmetry breaking scale and $M_Y$ is a heavy mass 
which should be of order $M_X$ if the operators have a field theoretic origin,
or $M_S$ if the operator's origin is from the string. 

 The $U(1)_X$ symmetry used is anomalous~\footnote{In fact, in the context of string
derived models, one linear combination among the various $U(1)$ symmetries accompanying 
the non-abelian gauge group factors, is anomalous\cite{aehn,alr,af}.}.
However,  the $U(1)$ charges  $Q_X$ are such that 
all of the mixed anomalies are equal, so that   we can appeal to Green-Schwarz 
string anomaly cancellation. Moreover, the spontaneous breaking of the $U(1)_X$
via the $\theta,\bar\theta$-VEVs at a high scale should proceed along a
 flat direction to avoid large vacuum energy contributions from D-terms.

 Using the $\theta,\bar\theta$ fields,
 we generate all effective Yukawa terms (except the (33) elements) by 
operators of the form
\begin{equation}
O_{ij}\sim (\bar{F}_i F_j)h\left(\frac{H\bar{H}}{{M_Y}^2}\right)^r
\left( \frac{\theta^n \bar{\theta}^m}{{M_S}^{n+m}} \right)
\sim 
\bar{F_i} F_j h \delta^r \epsilon^{n+m},
 \label{op}
\end{equation}
where $r,n,m \ge 0$ are set so that the operator
is invariant under the U(1)$_X$ family symmetry.
The (33) operator is kept to be of the renormalisable (trilinear) form.
When $H$ and
$\bar{H}$ develop their VEVs, the operators in eq. (14) will become effective
Yukawa couplings of the form $\bar{F} F h$ with a small coefficient of
order $(M_{GUT}^2/M_{Y}^2)^r$.  
\begin{table} 
\begin{center}
\begin{tabular}{ccccc} \hline
 & $Q \bar{U} h_2$ & $Q \bar{D} h_1$ & $L \bar{E} h_1$ & $L \bar{N}
h_2$
\\ \hline
$O^A$ &1 & 1 & 1 &1 \\ 
$O^C$ &$\frac{1}{\sqrt{5}}$ & $\frac{1}{\sqrt{5}}$ 
& $\frac{-3}{\sqrt{5}}$ &$\frac{-3}{\sqrt{5}}$  \\ 
$O^M$ &0 & $\sqrt{2}$ & $\sqrt{2}$ &0 \\ 
$O^W$ &0 & $\sqrt{\frac{2}{5}}$& -3$\sqrt{\frac{2}{5}}$&0\\
\hline
\end{tabular}
\end{center}
\label{tab:subset}
\caption{{\protect\small When the Higgs fields develop their VEVs at $M_{GUT}$, the
$r=1$ operators utilised lead to the effective Yukawa couplings with
Clebsch - Gordon coefficients as shown. We have included the relative
normalisation for each of the operators.
}}
\end{table}
The see-saw mechanism for suppressing the 
mass of the three lighter neutrinos is
realized in a natural way,
provided that non-renormalisable terms of the form
\beq
\bar{F}_i\bar{F}_j\left( \frac{H H}{M_{X}} \right) 
\left( \frac{\theta^o \bar{\theta}^p}{{M_S}^{o+p}} \right)
\sim \bar{F}_i\bar{F}_j  M_{GUT} \sqrt{\delta} \epsilon^{o+p}
\label{maj1}
\eeq
are included in the superpotential.  
It should be noted that when the exponents in
eq.~(\ref{maj1}) are not integers, the corresponding element is zero due to a
residual discrete symmetry. Gauge unification occurs at $M_{GUT}$, and
so the gauge couplings must run together between $M_{GUT}$ and ${M_S}$.
This is possible, for example, when some extra states are present in the 
above range [12].
\section{Textures \label{asymmetric}}
 
  In ref.~\cite{AKLL}, we showed that textures exist which fit 
the data {\em and}
\/explain the presence of the zeroes through highly suppressed operators.
The family mass hierarchy itself is explained through suppression via U(1)$_X$
family symmetry: all of the fundamental Yukawa couplings are of order 1 in the
model.
Rather than perform a complicated systematic analysis of all possible models,
here we simply examine  the predictions for the neutrino masses and the lepton mixing 
of one successful model.  
We choose an example of a texture (model 1 of ref.~\cite{AKLL}), that
provides a very good quality of the fit of the data on charged fermion masses 
($\chi^2/\mbox{d.o.f.}=0.34$ for 3 d.f.). 
The texture is
\begin{equation}
\lambda = \left(\begin{array}{ccc}
0 & O^M+s.d. & 0 \\
O^M+O^A & O^W+s.d. & 0 \\
0  & O^C  & O_{33} \\ \end{array}\right) 
\label{nonsym}
\end{equation}
where $O_{33}$ is the renormalisable operator and ${s.d.}$ stands for small
sub-dominant operators which are negligible in the down quark and charged
lepton sectors, but which are responsible for the charm and up quark masses. 
The C-G structure of the operators $O^{A,C,M,W}$ is listed in
Table 1. 
They are arrived at by taking different contractions
of the SU(4)$\otimes$SU(2)$_R\otimes$SU(2)$_L$ group indices~\cite{AKLL}.
Operators listed in the table are the ones which were found to be useful
phenomenologically, and which are utilised in this paper.
The U(1)$_X$ symmetry is responsible for high powers of $\epsilon$,
reproducing the approximate zeroes assumed in eq.~(\ref{nonsym}).
We pick an assignment of $U(1)_X$ charges corresponding to family dependent
charges of case 2 of
ref.~\cite{AKLL}. 
The charge assignment for this case appears 
in Table~\ref{table:2}, where we have added family independent components to
the U(1)$_X$ charges in order to make the mixed anomalies of $SU(4)^2 U(1)_X$,
$SU(2)_R^2 U(1)_X$ and $SU(2)_L^2 U(1)_X$ equal. This allows anomaly
cancellation under the GSW mechanism.
\begin{table}[h]
\centering
\begin{tabular}{cccccccccc}\hline
 $F_1$ & $F_2$ & $F_3$ &  $\bar{F}_1$ & $\bar{F}_2$ & $\bar{F}_3$ & $h$ &
 $H$ & $\bar{H}$ \\ \hline
4 & 0 & -1 & 3 & -1 & 1 & 0 &  1 & -1 \\
\hline
\end{tabular}
\caption{{\protect\small $U(1)_{X}$ charges of fields in the model.}
\label{table:2}}
\end{table}
The operators resulting from the charge
assignments in Table~\ref{table:2} reproduce the mass/mixing hierarchies with
order 1 Yukawa couplings in
the charged fermion sector if $\epsilon=0.14$, $\delta=0.21$. We shall take
these values of expansion parameters for the analysis of the neutrino masses
below. If we take only the field content of Table~\ref{table:2}, 
the D-flatness constraint 
sets the natural scale of $|<\theta>|^2$ and
$|<\bar{\theta}>|^2$ to be of order $g_u^2 M_{Pl}^2 / (4 \pi^2)$, 
$g_u$ being the unified gauge coupling constant at $M_S$.

   At $M_{GUT}$, the Yukawa matrices are of the form
\begin{equation}
\lambda^I = \left[ \begin{array}{ccc} 0 & H_{12} e^{i \phi_{12}}
x_{12}^I 
& 0
\\ H_{21} x_{21}^I e^{i \phi_{21}}+ \tilde{H}_{21} \tilde{x}_{21}^I
e^{i\tilde{\phi}_{21}} & H_{22} x_{22}^I e^{i \phi_{22}} &
0 \\
0 & H_{32} x_{32}^I e^{i \phi_{32}} & H_{33} e^{i \phi_{33}} \\
\end{array}\right], \label{dom}
\end{equation}
where only the dominant operators are listed. The $I$ superscript
labels the charge sector, $x_{ij}^I$ refers to the Clebsch-Gordon
coefficient relevant to the charge sector $I$ in the $ij^{th}$
position, $\phi_{ij}$ are unknown phases and $H_{ij}$ is the magnitude
of the effective dimensionless Yukawa coupling in the $ij^{th}$
position. In our case $H_{ij}$ represent the effective 
small Yukawa couplings generated
by operators from eq.~(\ref{op}).
Any sub-dominant operators that we introduce will be denoted
below by a prime and it should be borne in mind that these will only
affect the up-quark and Dirac neutrino matrices. 
Once unphysical phases have been rotated away by re-phasing the fermion fields,
only one physical phase remains.
The charged fermion
analysis did not significantly constrain the remaining physical phase
$\phi$ and for
our purposes it shall be set to zero.

  Table 1 displays the Clebsch-Gordon coefficients derived from
the operators utilised for this model, a subset of operators used for this and
other models in ref.~\cite{AKLL}.
Putting in the Clebsch - Gordon coefficients from 
Table 1 we arrive at the 
component Yukawa matrices, at the GUT scale of
\begin{eqnarray}
\lambda^U &=& \left(\begin{array}{ccc}
0 & \eta_{12}  & 0 \\
H_{21}' &  H_{22}' & 0 \\
0  & \sqrt{2}H_{32}/\sqrt{5}   & H_{33} \\ \end{array}\right) 
\label{nonsymupcomponents}
\\
\lambda^D &=& \left(\begin{array}{ccc}
0 & \sqrt{2}H_{12} & 0\\
\sqrt{2}H_{21} & H_{22}/\sqrt{5} & 0 \\
0  & -\sqrt{2}H_{32}/\sqrt{5}    & H_{33} \\ \end{array}\right) 
\label{nonsymdowncomponents}
\\
\lambda^E &=& \left(\begin{array}{ccc}
0 & \sqrt{2}H_{12} & 0\\
\sqrt{2}H_{21} & 3H_{22}/\sqrt{5} & 0 \\
0  & -3 \sqrt{2}H_{32}/\sqrt{5}    & H_{33} \\ \end{array}\right) 
\label{nonsymleptoncomponents}
\\
\lambda^{\nu}_{Dirac} &=& \left(\begin{array}{ccc}
0 & \eta_{12}  & 0 \\
H_{21}' &  H_{22}' & 0 \\
0  & -3\sqrt{2}H_{32}/\sqrt{5}   & H_{33} \\ \end{array}\right) 
\label{neutmas}
\end{eqnarray}
where $ \eta_{12}$ is expected to be generated from a higher
order non-renormalisable  term.
 
The following GUT scale input parameters minimised the fermion mass/mixing
global $\chi^2$ for the case considered here~\cite{AKLL}:
\begin{eqnarray}
H_{22} = 2.88\times 10^{-2}, & H_{12} = 2.81\times 10^{-3}, &
H_{21} = 1.30 \times 10^{-3}, \nonumber
\\ H_{33} = 1.18,&  H_{22}' = 1.91 \times 10^{-3},&
H_{21}' = 1.94\times 10^{-3}. \nonumber \\
H_{32} = 7.28\times 10^{-2}, & \eta_{12} \cong 10^{-3}. &
\label{inputs}
\end{eqnarray}
\noindent These values of the parameters lead to the predictions
\begin{eqnarray}
\alpha_S(M_Z) = 0.119, & \tan \beta = 59.5, & m_t = 175,\nonumber \\
m_d = 6.25, & m_s = 158, & m_b = 4.24, \nonumber \\
m_c = 1.30, & |V_{us}| = 0.2211, & |V_{ub}|=3.71 \times 10^{-3}, \nonumber \\
m_u = 5, & |V_{cb}| = 0.038, &
\label{outputs}
\end{eqnarray}
where all masses are quoted in the $\overline{MS}$ renormalisation scheme, 
$m_{c,b,t}$ are running masses quoted in GeV and $m_{u,d,s}$ are evaluated at 1
GeV and measured in units of MeV. The predictions for 
$|V_{us}|$, $|V_{ub}|$ and $|V_{cb}|$ are at the scale $M_Z$. 
\section{Neutrino masses}

 Up to now, we have 
shown that our ansatz generates successfully the quark and
charged lepton mass hierarchy. In addition, the entries of the Dirac neutrino
mass matrix are completely determined at the unification scale since they
are related to those of the up-quark mass matrix: 
\begin{equation}
m_D(M_{GUT}) = \lambda^{\nu}_{Dirac}\upsilon_2
\end{equation}
Thus, the overall scale of the matrix is determined by the Higgs field  VEV which gives
mass to the up quark, $\upsilon_2=\langle h_2\rangle$,  while  the Dirac Yukawa
couplings  can be encoded in terms of the parameters $\delta, \epsilon$ as can 
 be seen from~(\ref{neutmas}).

  We turn now to the discussion of the heavy right-handed (RH) neutrino Majorana 
mass matrix $M_R$. Since the RH neutrinos are members of the multiplets
$\bar{F}^i_{x\alpha}$ (eq.~(\ref{RH})) accommodating right-handed quarks and
leptons, their charges are already fixed.  Thus, within a given charge
assignment, we have no freedom of choosing the relative orders of magnitude
for the elements of $M_R$. 
The $U(1)_X$ charge of $H$ (opposite to the charge of $\bar{H}$) was not
fixed by the charged fermion analysis. 
Once the $U(1)_X$ charges are chosen, the structure of $M_R$ is directly
determined by the
non-renormalisable terms of eq.~(\ref{maj1}). Therefore, the $M_R$ entries 
are also encoded in terms of  the same parameters $\delta$  and $\epsilon$
up to Yukawa coefficients of order 1. On the other hand,
the overall scale is related to the unification mass $M_{GUT}$. 
When the charges from Table~\ref{table:2} are substituted into
eq. (\ref{maj1}),   
it is found that the effective RH neutrino Majorana mass matrix is
\begin{equation}
M_R= M_{GUT} \sqrt{\delta} \left(
\begin{array}{ccc}
a\cdot \epsilon^{8} &\epsilon^4 & d \cdot \epsilon^6\\
\epsilon^4 & b & e \cdot \epsilon^2\\
d\cdot \epsilon^6 & e \cdot \epsilon^2 &c \cdot \epsilon^4\\
\end{array} \right) \label{maj}
\end{equation}
 In  eq. (\ref{maj}),  $a,b,c,d,e$ are the dimensionless Yukawa couplings,  
mentioned above.
In the case of the string derived version of the model, they are calculable
and determined in terms of the gauge coupling at the unification 
 scale. In the field theory version, however, 
they cannot be  determined,
so they will be treated as free real parameters.  
The suppression factors
appearing in the  heavy RH neutrino Majorana mass matrix 
will finally give $M_R < M_{GUT}$.
This implies that the  light neutrino masses which will be obtained
through the see-saw mechanism, will  turn
out to be larger than might be naively expected for $M_R \sim
M_{GUT}$.

In the following,  we will show that four of the five  hybrid solutions
found in ref. \cite{LP97} can be obtained for natural values of $a,b,c,d,e$.
We will describe first the procedure of our analysis. 
As has already been explained, the neutrino Dirac mass matrix is completely set   
by the charged fermion data. A decisive role is played by the involved
Yukawa couplings at the GUT scale. Since these are related to the up-quark
couplings, they are  obtained by using the 3-loop QCD$\otimes$1-loop QED-$\beta$
functions below $m_t$, and the 1-loop MSSM
$\beta$-functions between $m_t$ and $M_{GUT}$, as in ref. \cite{AKLL}. The
renormalisation effect of the Dirac Yukawa couplings of the neutrinos is
negligible because they are integrated out at the heavy Majorana mass scale. 
Only the $\tau$-neutrino Dirac coupling is large enough to affect the results,
and this is integrated out within a couple of orders of magnitude below 
$M_{GUT}$, i.e, at the scale where the see-saw mechanism is expected to
operate.
The effective light  left-handed (LH) neutrino Majorana mass matrix 
at $M_{GUT}$ is then 
$$m_{\nu_L}^{eff}\equiv - m_D^T M^{-1} m_D$$ 
The  light neutrino masses are  given by 
\begin{equation}
{\cal O} m_{\nu_L}^{eff}
 {\cal O}^T = \left( \begin{array}{ccc} m_1 & 0 & 0 \\
    0 & m_2 & 0 \\
    0 & 0 & m_3 \\ \end{array} \right),
\end{equation}
where ${\cal O}$ is an orthogonal matrix that diagonalises
$m_{\nu_L}^{eff}$ and $m_{i}$ are the light neutrino 
masses ($m_1 \ll m_2 \ll m_3 $). 
   The lepton mixing matrix is calculated by
$U={\cal O} V_{E_L}$, where the matrix $V_{E_L}$ 
appears as a result of the diagonalisation of the charged lepton mass matrix.
The mixing elements and $m_{1,2,3}$ are then
renormalised from $M_{GUT}$ down to $m_t$, again using the 1-loop MSSM
RGEs~\cite{AK46}. The renormalisation of mixing angles and masses between $m_t$
and $M_Z$ is small and  neglected.
\begin{figure}
\begin{center}
\threegraphs{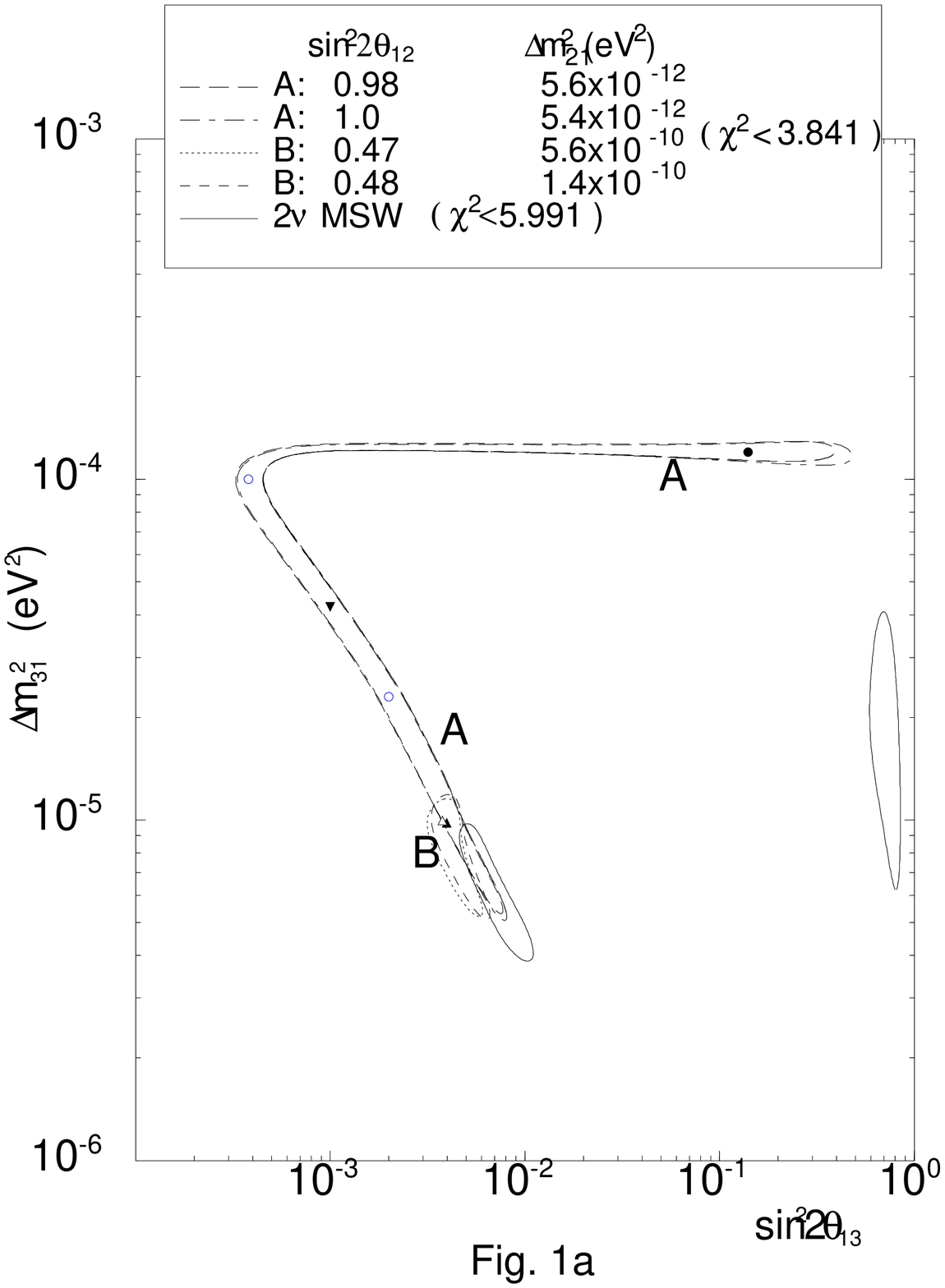}{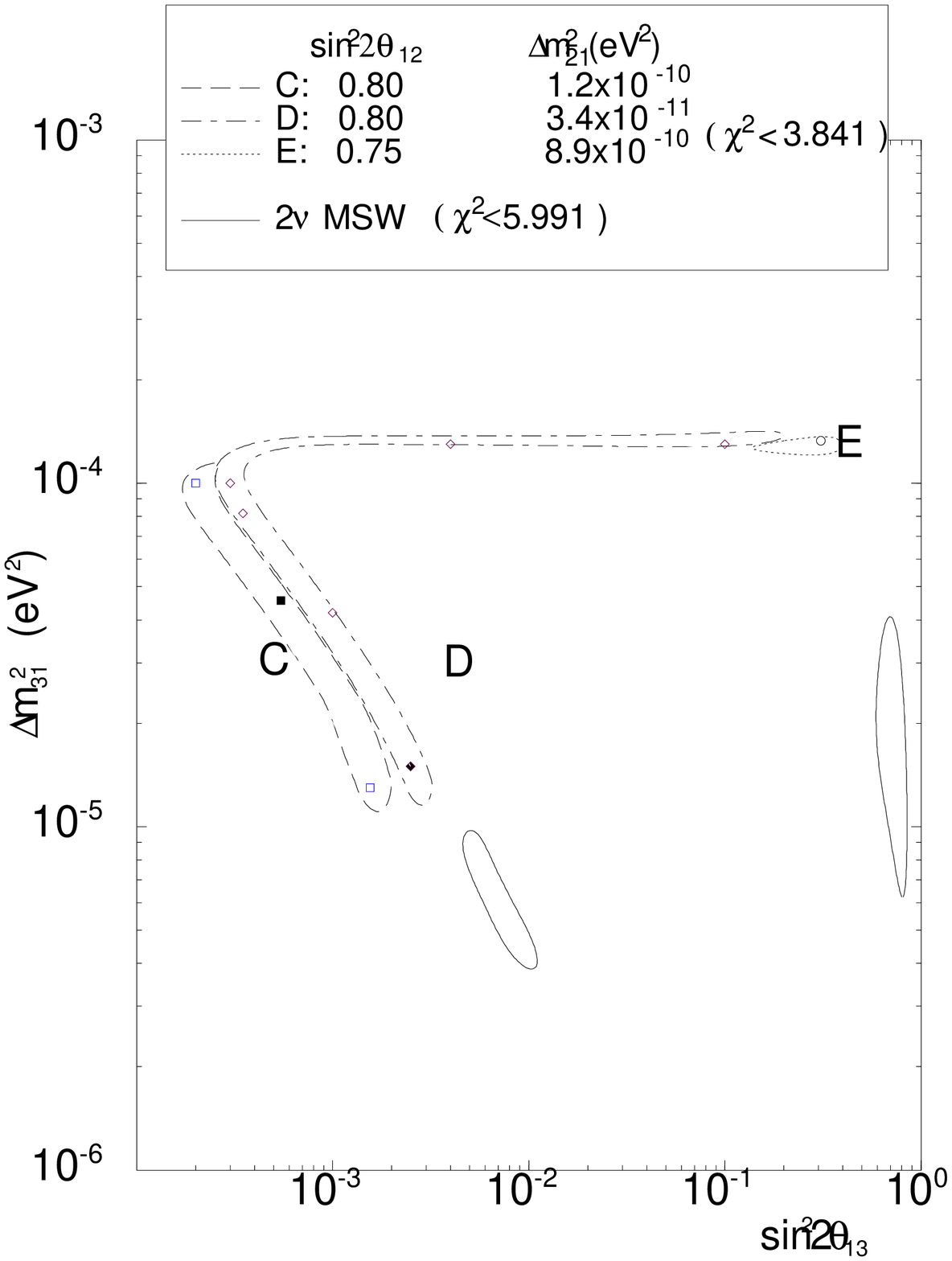}{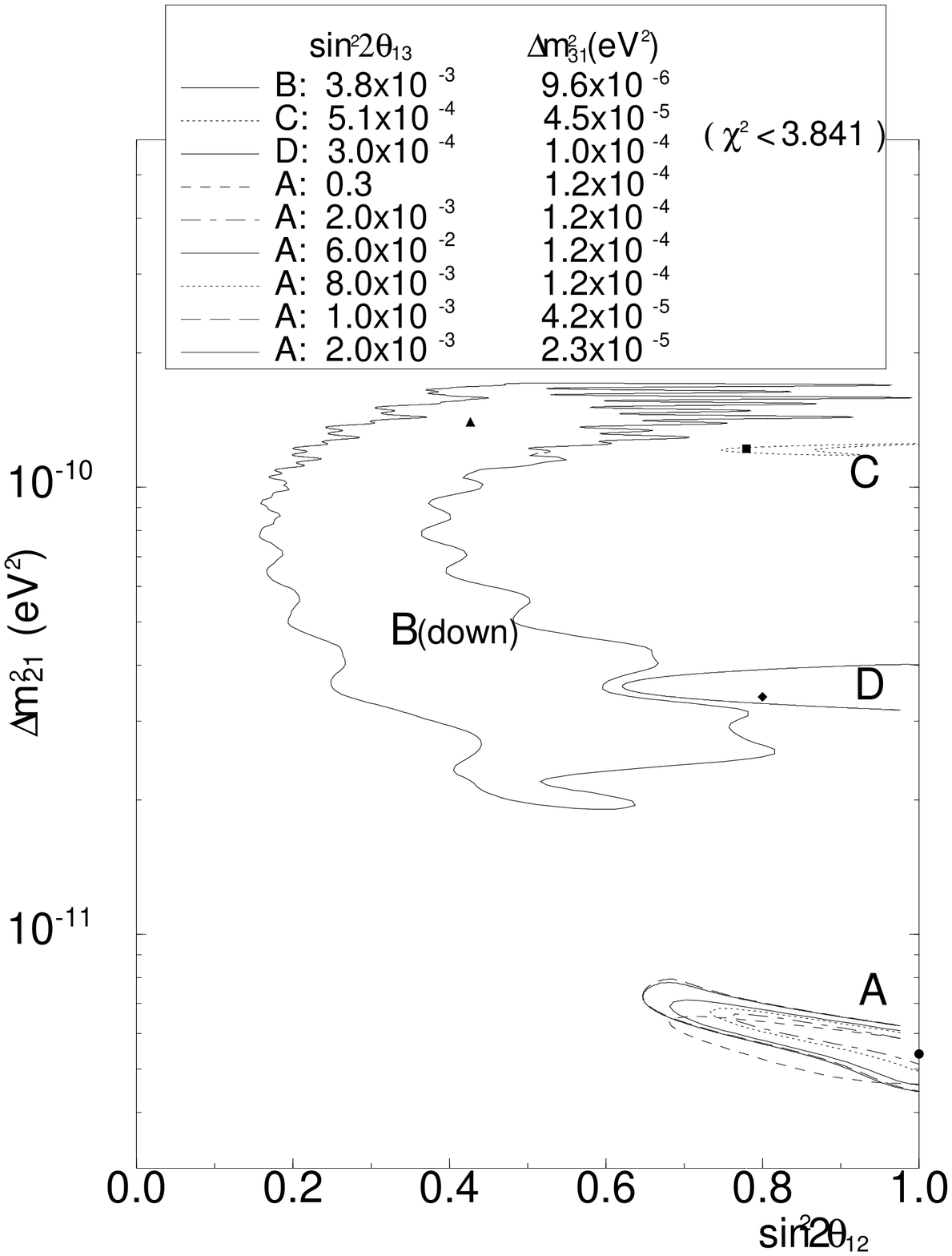}
\end{center}
\caption{MSW + VO solutions of the solar neutrino problem~\protect\cite{LP97} -
 regions of values of the MSW transition and vacuum oscillation parameters.
 (a) and (b) show $\Delta m_{31}^2$
  and $\sin^2 2\theta_{13}$, corresponding to solutions A,B,C,D and E and
  obtained 
  for fixed values of the vacuum oscillation parameters $\Delta m_{21}^2$
  and $\sin^2 2\theta_{12}$ indicated in the figures. Shown are also
  the 95\% C.L. $2\nu$ MSW non-adiabatic and adiabatic solution regions.
 (c) shows $\Delta m_{21}^2$ 
  and $\sin^2 2\theta_{12}$ for solutions A, B, C, and D.
  The regions correspond to fixed values of
  the MSW parameters $\Delta m_{31}^2$
  and $\sin^2 2\theta_{13}$ indicated in the figure.}
\label{fig:LP}
\end{figure}
%
\begin{table}
\centering
\begin{tabular}{cccccccccc}\hline
 &a&b&c&d&e&$\Delta m_{31}^2$&$\Delta m_{21}^2$&$\sin^2
 2\theta_{13}$&$\sin^2
2\theta_{12}$ \\
\hline
A&1.0&0.25&0.11&0.125&0.25&
$1.2\ 10^{-4}$ &$ 5.5 \ 10^{-12}$ &$ 2.5 \ 10^{-3}$ & 1.00 \\
B&0.75&0.625&0.407&1.2&0.457&9.6\ $10^{-6}$&8.7\ $10^{-11}$ & 
3.8\ $10^{-3}$ & 0.25\\
C & 1.0 & 0.345 & 0.7 & 2.0 & 0.3 & 
1.4\ $10^{-5}$ & 1.2\ $10^{-10}$ & 1.6\ $10^{-3}$ & 0.80\\
D & 1.7 & 0.25 & 0.24 & 1.91 & 0.4 &
1 \ $10^{-4}$ & 3.9\ $10^{-11}$ & $3.0\ 10^{-4}$ & 1.00\\
\hline
\end{tabular}
\caption{{\protect\small Examples of values of
$\Delta m_{31}^2$, $\Delta m_{21}^2$, $\sin^2 2\theta_{13}$ and $\sin^2
2\theta_{12}$ obtained in the model discussed by us and
corresponding to the hybrid MSW + VO solutions
A, B, C and D found in ref. \protect\cite{LP97}
($\Delta m_{ij}^2$ are in units of eV$^2$). The dimensionless real parameters
a,b,c,d and e from the heavy RH neutrino Majorana mass matrix are supposed to 
take values in the interval (0.1 - 2.0).}}
\label{tab:hybsolns}
\end{table}

Our string motivated criteria for the naturalness of the Yukawa couplings
$a,b,c,d,e$, is that they lie in the range $0.1-2.0$.
By varying $a,b,c,d,e$, four predictions corresponding to hybrid oscillation
solutions A, B, C, D were found that were consistent with this constraint.
These are summarized  in Table~\ref{tab:hybsolns} and should be compared
with the regions shown in Figs.~\ref{fig:LP}a,~\ref{fig:LP}b and~\ref{fig:LP}c.

The solutions in Table~\ref{tab:hybsolns} correspond to the following 
$M_{GUT}$, masses of the 
heavy (approximately RH) neutrinos $M^D_{Rj}$, $j=1,2,3$, 
and lepton mixing matrices 
(defined in the basis $(\nu_1, \nu_2, \nu_3)$,  $m_1 < m_2 < m_3$): 
\begin{eqnarray}
{\mathrm Solution~A}: &&M_{GUT}= 0.9 \times 10^{16},\, M^{D}_R = \left( 6.8 \times 10^{14}, 
4.1 \times 10^{7}, 3.0 \times 10^{13} \right) \nonumber \\ &&
U^A \cong
\left( \begin{array}{ccc}
0.69    & -0.72      & -0.025       \\
-0.72   &  -0.69     & 0.038       \\
-0.045    &   -0.0080    &   -1.00     \\
\end{array} \right) \\
{\mathrm Solution~B}:&&
M_{GUT}= 0.8 \times 10^{16},\,
M^{D}_R = \left( 2.9 \times 10^{15}, 
7.7 \times 10^{7}, 1.5 \times 10^{13} \right)
\nonumber \\
&&U^B \cong
\left( \begin{array}{ccc}
-0.26    & 0.96      &  0.031      \\
0.96    & 0.26      &  0.057      \\
0.047    &  0.044     & - 1.00      \\
\end{array} \right) \\
{\mathrm Solution~C}:&&
M_{GUT}= 2.5 \times 10^{16},\,
M^{D}_R = \left( 1.4 \times 10^{15}, 3.0 \times 10^{7}, 7.2 \times 10^{12} \right)
\nonumber \\
&&U^C \cong
\left( \begin{array}{ccc}
-0.52    & 0.85      & 0.020       \\
0.85    &   0.52    &  0.066      \\
0.046    &   0.052    & - 1.00      \\
\end{array} \right) \\
{\mathrm Solution~D}:&& M_{GUT}= 1.9 \times 10^{16},\,
M^{D}_R = \left( 6.8 \times 10^{14}, 1.4 \times 10^{7}, 1.8 \times 10^{13} \right)
\nonumber \\
 &&U^D \cong
\left( \begin{array}{ccc}
-0.69    &  0.72     &  -0.0086      \\
0.72    & 0.69      &  0.062      \\
0.051    &  0.037     &  -1.00      \\
\end{array} \right) 
\end{eqnarray}
\noindent where the $M_{GUT,R}$  masses are in GeV.

 The MSW + VO solutions A, B, C, and D of the solar neutrino problem
of interest will be tested in the currently operating 
Super-Kamiokande experiment as well as in the
future solar neutrino experiments SNO, BOREXINO, HELLAZ, etc.
As discussed in \cite{LP97}, one of the distinctive
predictions of the indicated hybrid MSW + VO solutions 
is the existence of strong and very characteristic distortions of
the spectrum of $^{8}$B neutrinos, which should be observable 
in the Super-Kamiokande and/or SNO experiments. 
The seasonal (time) variation of the $^{8}$B, $^{7}$Be, pp, etc.\ solar neutrino
fluxes caused by\footnote{The solar neutrino flux at the Earth surface 
changes with the time of the year due to the ellipticity
of the Earth orbit around the Sun. The change of the flux from December to June 
due to the standard geometrical effect is 6.68{\%}.}
the vacuum oscillations (see, e.g., \cite{BP78} and the references quoted therein)
in the case of the MSW + VO solutions, cannot exceed 
for given values of
$\Delta m_{21}^2$ and $\sin^2 2\theta_{12}$,  
the seasonal variation caused by purely two-neutrino
vacuum oscillations with the same
$\Delta m_{21}^2$ and $\sin^2 2\theta_{12}$~\footnote{This result follows directly from eq. (9) or eq. (17) in ref. \cite{LP97}.}. 
 Nevertheless, the vacuum oscillation induced seasonal variations of 
the $^{7}$Be and  $^{8}$B neutrino fluxes predicted for the
MSW +VO solutions B and C,
can be observable in the Super-Kamiokande, SNO and BOREXINO experiments
for certain regions of the solution values of the parameters 
\cite{LP97}. For solutions A and 
D this effect will be observable for 
the $^{7}$Be and $pp$-neutrinos \cite{KP96}, if the $pp$-neutrino flux
is measured with detectors like HELLAZ or HERON\@.
Finally, for the MSW + VO solution parameters we have obtained in the model
studied here, the day-night asymmetry caused by $\nu_e$ and $\nu_{\mu (\tau)}$ 
MSW transitions in the Earth is estimated to be rather small  
in the signals of the detectors sensitive only to $^{8}$B or 
$^{7}$Be neutrinos \cite{LP97, MP97}. 
\vskip -0.2truecm
\section{Conclusions}
\vskip -0.2truecm
In the present paper we have investigated the possibility to
accommodate the hybrid MSW + VO solutions of  the
solar neutrino problem, found recently in \cite{LP97},
in a string motivated partially unified model.
The model is based on the $SU(4)\otimes O(4)$ gauge group and was shown
to  provide successful predictions in the charged fermion
mass sector and  fit the data well. In addition, a
U(1)$_X$ family symmetry allows the generation of effective Yukawa
couplings which are naturally small, without having to put very small
fundamental Yukawa couplings in the model by hand. The charged fermion
data fit constrains the Dirac neutrino mass matrix and the magnitudes of
the elements of the RH-neutrino Majorana mass matrix are then set
by the $U(1)_X$ family symmetry.
 
 We have shown in the present work that 
the new hybrid MSW + VO solutions to the
solar neutrino problem, found in \cite{LP97}, appear naturally 
in this rather compelling model. 
The horizontal $U(1)_X$ symmetry plays a vital role in suppressing certain
elements of the RH-neutrino Majorana mass matrix, allowing some of the
light neutrinos to be heavier than one would be naively expecting
and therefore to lie in the mass ranges required.
The vertical symmetry relates the
Dirac neutrino masses to the up quark masses, allowing more
predictivity.
This is evident in the fact that the whole $3\times 3$ lepton
mixing matrix is unambiguously determined. 
The predictions of the model for the neutrino masses and the lepton mixing
will be tested in the currently operating and the future solar 
neutrino experiments (Super-Kamiokande, SNO, BOREXINO, HELLAZ, etc.).      
\vskip -0.2truecm
\section*{Acknowledgements}
\vskip -0.2truecm
B.C.A. would like to thank CERN for hospitality while part of this work was
carried out. The work of G.K.L. is
partially supported by TMR-ERBFMXR-CT96-0090.
The work of S.T.P. was supported in part
by the EEC grant ERBFMRX-CT96-0090 and by Grant PH-510 from the
Bulgarian Science Foundation.

 
\end{document}